# Real time simulation of 2007 Santa Ana fires

A. K. Kochanski[1], M.A. Jenkins[2], S. K. Krueger[1], J. Mandel[3] and J. D. Beezley[3]
[1]*Dept of Atmospheric Sciences University of Utah, Salt Lake City, UT,*
[2]*Department of Earth and Space Science and Engineering, Faculty of Pure and Applied Science York University, Toronto, Ontario, Canada,*
[3]*Department of Mathematical and Statistical Sciences, University of Colorado, Denver*
[adam.kochanski@utah.edu](adam.kochanski@utah.edu)

## Abstract
There are many wildfire behaviors of increasing relevance that are outside the forecast capabilities of even the most sophisticated operational fire spread and fire behavior model. The limitations of the operational models are due primarily to their inability to represent coupled fire-atmosphere interactions. Coupled wildfire-atmosphere models are physics-based fluid-dynamical prognostic models of wildfire spread and behavior that attempt an almost complete representation of fire-atmosphere interactions. This level of fidelity however means that these models cannot be used operationally. The reason is that, despite ever increasing computational resources, the complexity and range of processes and scales (1 mm to 100 km) involved in this modeling approach make computational costs prohibitively expensive. In this study we propose an intermediate approach. A physics-based coupled atmosphere-fire model is used to resolve the large-scale and local weather as well as the atmosphere-fire interactions, while combustion is represented simply using an existing operational surface fire behavior model. This model combination strikes a balance between fidelity and speed of execution. The feasibility of this approach is examined based on an analysis of a numerical simulation of two very large Santa Ana fires using WRF-Sfire, a coupled atmosphere-fire model available at the Open Wildland Fire Modeling Community (OpenWFM.org); an earlier version is available as WRF-Fire in WRF release. The study demonstrates that a wind and fire spread forecast of reasonable accuracy was obtained at an execution speed that would have made real-time wildfire forecasting of this event possible.

## 1. Introduction
There are multiple simulator models for operational forecasting of forest fire propagation as shown by Papadopoulos and Pavlidou (2011), and Sullivan (2009), who examine each simulator in turn, discussing their attributes and capabilities, along with their drawbacks and deficiencies. The conclusion of both studies is that, of the existing simulators, FARSITE is the most precise. To ensure the best forecast, FARSITE necessitates ingestion of multiple layers of data. Spatially-gridded GIS observational data on fuels and topography are required, and weather data are required to provide surface wind speed and direction, temperature, humidity, and cloud cover at time of ignition. The primary end product of FARSITE is the prediction of a fire perimeter over the fire's landscape. Sullivan (2009) points out that current operational fire-spread models are a conversion of one-dimensional linear models of fire spread to two-dimensional models of fire spread, and FARSITE is no exception. FARSITE is based on BEHAVE (Andrews 1986) which is based on the rate-of-spread (ROS) model by Rothermel (1972).

Despite its wide-spread use in the United States and elsewhere, FARSITE, along with the other operational fire-spread formulations discussed by Papadopoulos and Pavlidou (2011), and Sullivan (2009), suffer from one fundamental defect, and that is their simplistic treatment of the wind on fire behavior. These models consider only surface wind direction and strength, they lack a real-time wind and weather forecast component, and they fail to account for coupled atmosphere/wildfire interactions.

In the scientific community there exists a significant number of physics-based fluid-dynamical deterministic numerical modeling studies (e.g., Mell et al. 2007, Colman and Linn 2007, Coen 2005, Sun et al. 2009, Mandel et al. 2011) demonstrating the significant impacts changing environmental wind conditions and coupled atmosphere/fire flow have on wildland fire propagation. Despite the physical validity of a fluid-dynamical coupled atmosphere/fire numerical model for predicting fire spread, operational application of this type of model is thought to be beyond present computing capabilities. The prevailing view in both scientific and operational communities is that wildfire behavior prediction using this modeling approach must therefore remain relegated to the study of wildfires under conditions not amenable to field experimentation.

In this study we demonstrate that this view may no be longer true. Using readily available computing capabilities, along with spatially-gridded GIS data on fuels and topography, we have chosen to use the WRF-Sfire (Mandel at al. 2009, 2011), a physically-based coupled atmosphere-fire modeling system, to simulate two wildland fires that burned during a Santa Ana weather event. Our overall objective is to test the feasibility of WRF-Sfire for accurate real-time forecasting of wildfire behavior. To achieve this objective, we perform a faster-than real-time simulation of two 2007 Santa Ana fires and compare the results to available weather and fire observations.

In this study we analyze the WRF-Sfire numerical simulation of Witch and Guejito fires which started on 21 October 2007 at 19:15 UTC and 22 October 2007 at 08:00 UTC respectively. They spread under strong Santa Ana winds burning 56,796 ha, leading to $18 millions in damage and two fatalities. Together they were the second largest fire event of the 2007 California wildfire season (Keeley et al. 2009).

The paper is organized as follows. We describe the WRF-Sfire and its forecasting abilities in Section 2. In Section 3 we present a WRF-Sfire model setup that allows for real-time weather and fire-spread prediction. Using this model configuration and setting the initial fuel and weather conditions based on data described in Section 3.2, the WRF-Sfire was run for each fire and final wildfire forecasts were produced. The accuracy of the model results are analyzed: first in terms of providing a realistic wind forecast, and second in terms of providing a realistic fire-spread forecast. These results are presented, respectively, in Sections 4.1 and 4.2, where we compare simulated to observed winds in the vicinity of the fires, and simulated to observed fire progression and final fire perimeters. The paper is summarized and conclusions are given in Section 5.

## 2. Model description

WRF-Sfire is a coupled atmosphere-fire model, available from the Open Wild Fire Modeling Community (OpenWFM.org). It combines the WRF (Weather Research and Forecasting system) (Skamarock et al. 2008) with fire propagation (Clark et al., 1996, 2004, Patton and Coen 2004) calculated by the level set method (Mandel et al. 2009). The two-dimensional surface propagation of the fire perimeter is modeled by the advection of the level set function by the local fire rate-of-spread (ROS). Sullivan (2009) would categorize the WRF-Sfire as a quasi-physical model; i.e., it includes the physics of the coupled fire/atmosphere but it does not attempt to represent the chemistry of fire spread. Instead, the ROS is computed based on local fuel properties, slope, and wind speed using the semi-empirical Rothermel fire spread model (Rothermel 1972). In this way, the computational costs of the WRF-Sfire remain reasonable while, unlike existing operational fire spread models, atmosphere/fire coupled WRF-Sfire winds at the fire line are used to compute the surface fire spread, amount of fuel burned, and propagation of the fire front. Coupling between the fire and the atmosphere occurs mainly through the release of the latent and sensible heat at the surface by the fire into the WRF model of the atmosphere. The rate and amount of latent and sensible heat released depend on the fire perimeter's ROS as it evolves in time. As a result, the model atmosphere "feels the fire" and responds to it by changing air temperature, density, humidity, pressure, and the local wind field. This coupled fire/atmosphere local wind affects the fire spread and its intensity, allowing for a constant feedback between the fire and atmosphere. Figure 1 illustrates the two-way fire-atmosphere coupling used in WRF-Sfire.

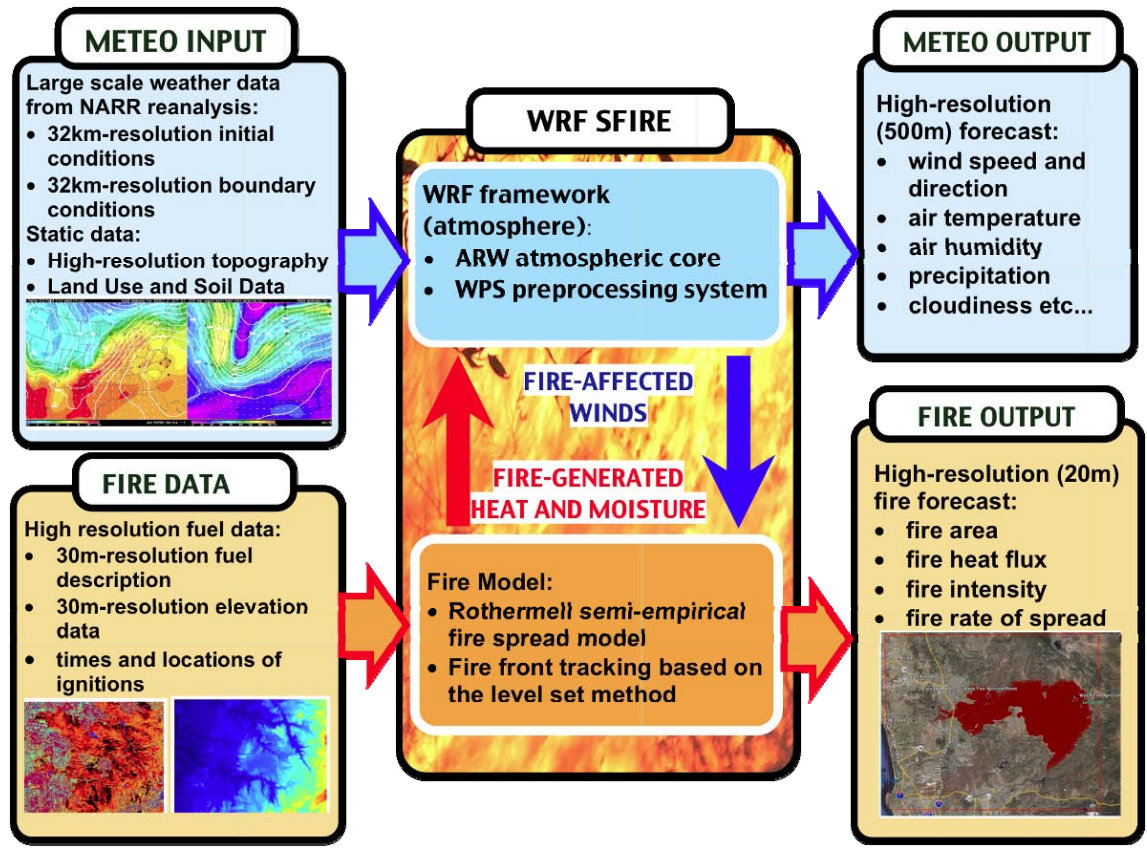

Figure 1. Diagram of the WRF-Sfire coupled fire-atmosphere modeling system.

The incorporation of WRF-Sfire into the WRF modeling framework (which is used for routine numerical weather prediction in the United States) allows for the use of detailed descriptions of the land use, fuel types, and topography (Beezley 2011, Beezley et al. 2011), and for realistic fire-atmosphere simulations that are affected by terrain and time-varying larger-scale meteorological forcing. The nesting capabilities of WRF (Figure 2) allow for multi-scale simulations in which a coarse tens-of-kilometers resolution outer domain captures the large synoptic-scale flow and feeds a set of nested higher-resolution domains. In this way, larger-scale to smaller- scale flows are gradually resolved to finally represent coupled atmosphere/fire flows at the smallest resolved scale. In addition complex terrain that influences small-scale flow is rendered more accurately. To accommodate high-resolution fuel and elevation data, and to provide sufficient accuracy for the fire spread computation without increasing computational cost, the fire model operates on a separate surface model grid refined significantly with respect to the atmospheric model (usually 15 to 25 times denser). Fire spread is therefore forecast at a resolution much finer than the resolution of the weather component of the WRF model. An example of the nested setup used for this study is presented in Figure 2. The nested setup and vertical grid refinement provide localized fire spread and weather predictions at significantly higher resolutions than currently available from NOAA (i.e., hundreds of meters resolution, versus 12 km resolution).

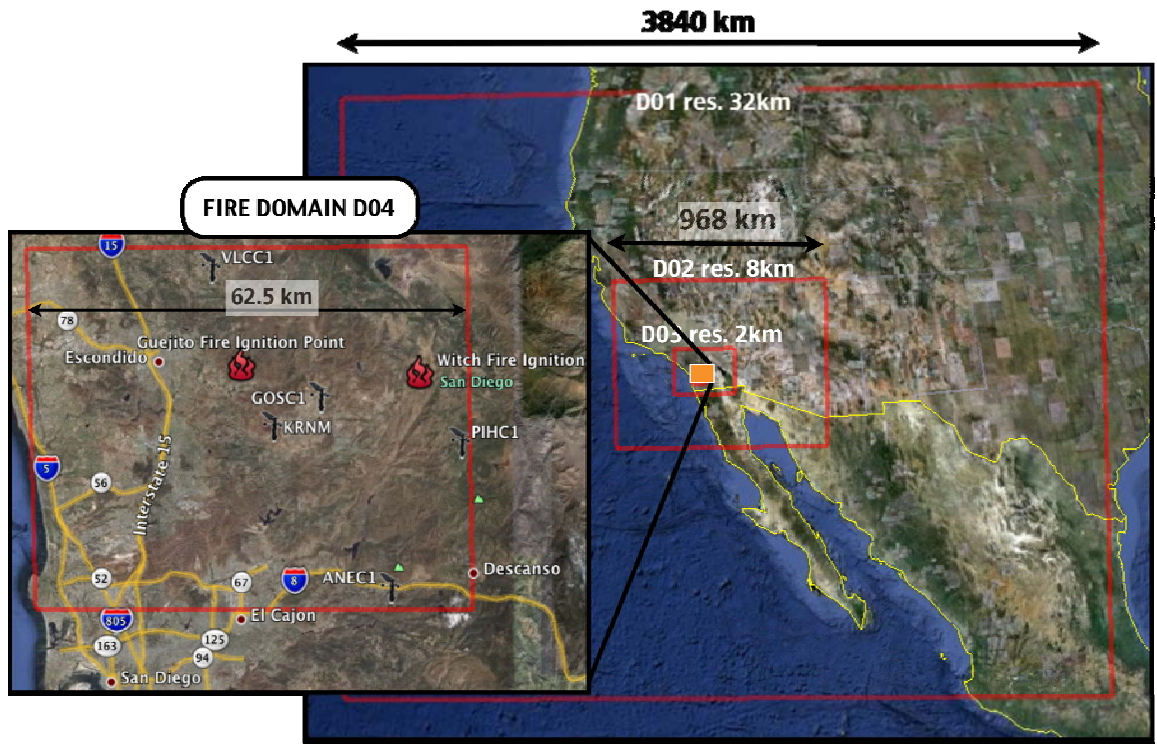

Figure 2. The multi-scale WRF setup used in this study with locations of fire origins and local meteorological stations used for model validation. Horizontal domain resolutions vary from 32km (D01) to 500m (D04).

## 3. Experimental setup

### 3.1. Model configuration

The Witch and Guejito fires simulated in this study were driven by strong westerly Santa Ana winds induced by a high-pressure system located over northern Nevada. As the pressure built up and the system moved eastward, southern California began to experience very strong and gusty Santa Ana winds that brought very warm and dry air from the Nevada desert into the San Diego area. In order to resolve the development and movement of this large-scale weather system together with the local circulation that is affected strongly by the complex topography of southern California, WRF was configured in the nested mode with four domains, D01, D02, D03, and D04, of horizontal-grid sizes 32km, 8km, 2km, and 500m, respectively. The domain setup used in this study is shown in Figure 2. A vertically-stretched grid was used to provide a high vertical resolution at the surface ($\Delta Z \sim 20m$), decreasing to coarser resolution ($\Delta Z \sim 500m$) between 3.5 and 7.5km altitude, and decreasing further ($\Delta Z \sim 2000m$) at the model top at 15.4km altitude. The fire domain was embedded within the finest domain (D04) of 500m resolution. The WRF-Sfire atmosphere-fire refinement ratio in the X and Y directions was set at 25, making 20m the horizontal fire-grid length. The details of the model configuration are presented in Table 1.

The 72h forecast presented in this study was run on 10 dual Intel Xeon X5670 nodes connected using QDR Infiniband links. Each node was equipped with two 6-core CPUs, so there was 12-cores available for each node. The entire 72h forecast was computed in 4h 48min, while the first 24h was ready in 1h 35min. The model output was saved at 10-minute intervals.

Table 1. Details of the WRF-Sfire setup.

| Domain | Atmospheric domain size X×Y×Z | Atmospheric horizontal resolution $\Delta X \times \Delta Y$ | Atmospheric vertical grid resolution $\Delta Z$ | Fire domain size $X_f \times Y_f$ | Fire domain resolution $\Delta X_f \times \Delta Y_f$ |
|---|---|---|---|---|---|
| D01 | 120×96×37 | 32km×32km | 20m-2000m | - | - |
| D02 | 121×97×37 | 8km×8km | 20m-2000m | - | - |
| D03 | 137×105×37 | 2km×2km | 20m-2000m | - | - |
| D04 | 125×105×37 | 500m×500m | 20m-2000m | 3125×2625 | 20m×20m |

### 3.2. Data sources

The WRF atmospheric component of the WRF-Sfire was initialized with the North American Regional Reanalysis (NARR) dataset. NARR surface and upper-air meteorological observations of wind, temperature, humidity, etc., provided the 3D description of the initial WRF-Sfire atmospheric state. The same data set was also used to create boundary conditions for the outer-most domain. The three outer domains of the WRF atmospheric component (i.e., D01, D02, D03) used MODIS-derived land-use categories and topography, while the innermost domain (D04) that contained the fire used 30m resolution National Elevation Dataset (NED) obtained from the USGS Seamless Data Warehouse (http://seamless.usgs.gov/), interpolated to 500m (Figure 3a).

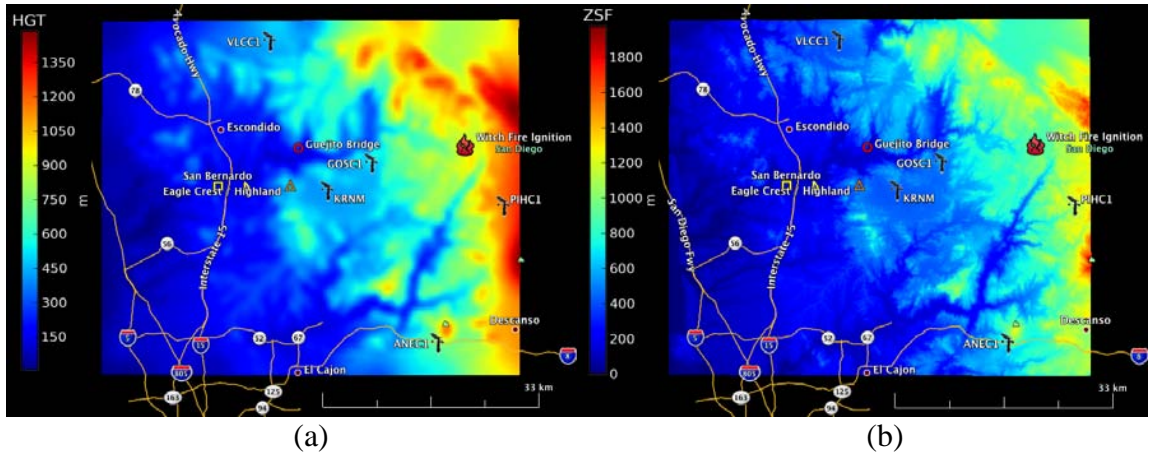

(a)          (b)

Figure 3. Elevation data used in the fire simulation (a) interpolated to the atmospheric grid of 500m resolution and (b) interpolated to the fire grid of 20m resolution.

The fire model was initialized with 30m resolution fuel data from LANDFIRE (Ryan et al 2006, Rollins 2009) in the form of 13 fuel categories as defined by Albini (1976). The special LANDFIRE categories not supported by the Rothermel model, like urban (91), snow/ice (92), and agriculture (93) were treated as missing data. The pixels marked as missing data were replaced with the prevailing categories of their surroundings, using the nearest-neighbor average option available in the WRF preprocessing system (WPS). Water (98) and barren lands (99) were converted into the no-fuel category (14). Even though the resolution of the fire data is very high, an analysis of the fuel maps revealed that some of the rivers (for example, the San Diego River) were not represented as continuous contours. Therefore the fuel map was compared with the visible-spectrum satellite image from Google Earth, and the discontinuities were fixed manually. The residential area of Ramona, represented in the fuel data as brush categories (5 and 6), was contoured as a no-fuel category to prevent fire from penetrating into the urban area. The unintended impact of this was to represent to some degree the fire suppression activities that were employed there. The final processed fuel map used for the fire simulation is presented in Figure 4. The topography for the fire model was obtained from NED and interpolated to 20m resolution (see Figure 3 b).

The version of WRF-Sfire used in this simulation does not support spatially- or temporally- varying ground fuel moisture content. To determine a single value to approximate fuel moisture within the fire domain, averages of temperature and humidity values recorded by 4 meteorological stations located within the computational domain (i.e., Ramona Airport- KRNM, Pine Hills - PIHC1, Valley Center - VLCC1, and Goose Valley - GOSC1) for 7h prior to the observed Witch fire ignition were calculated. Based on these calculations, the average fuel moisture content using Van Wagner (1969)'s model was estimated to be 6.6%.

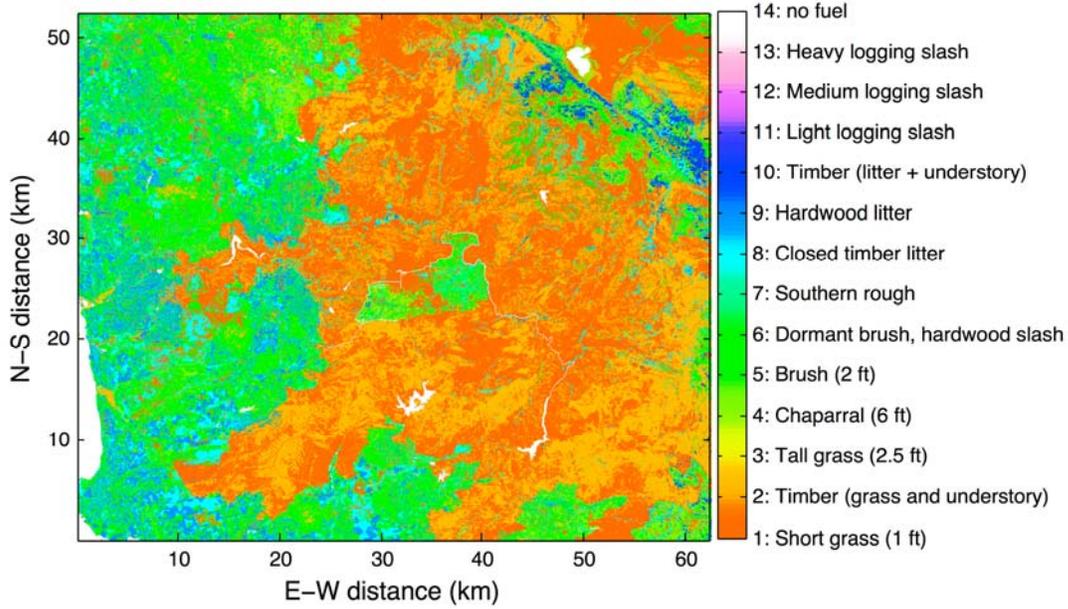

Figure 4. Fuel map used in the WRF-Sfire simulation according to the 13 Albini (1976) fuel categories.

To verify this value, the average fuel moisture for 8 dominant fuel categories within the fire domain was also estimated using the relative fuel loads derived from Albini (1979) and the fuel moisture content corresponding to low atmospheric moisture conditions as suggested by Scott and Burgan (2005). For each dominant fuel category we multiplied the relative contributions of each fuel class (1h, 10h and 100h) by its moisture content from Scott and Burgan (2005) (Table 2). An average of these dominant fuel categories was 6.58%. The final ground fuel moisture content for the WRF-SFIRE simulation was 6.5%, the average of these two estimates.

Table 2. Data used for ground fuel moisture estimates

| Fuel cat/class | Relative contributions of different fuel classes | | | Total per category (%) |
|---|---|---|---|---|
| | 1h | 10h | 100h | |
| 1 | 1.00 | 0.00 | 0.00 | 6.00 |
| 2 | 0.57 | 0.29 | 0.14 | 6.57 |
| 4 | 1.00 | 0.00 | 0.00 | 6.00 |
| 5 | 0.45 | 0.36 | 0.18 | 6.73 |
| 6 | 0.67 | 0.33 | 0.00 | 6.33 |
| 7 | 0.25 | 0.42 | 0.33 | 7.08 |
| 8 | 0.25 | 0.42 | 0.33 | 7.08 |
| 9 | 0.30 | 0.20 | 0.50 | 7.20 |
| moisture content (%) | 6.0 | 7.0 | 8.0 | **6.58** |

## 4. Model results and validation

In coupled atmosphere-fire simulations, the predicted fire behavior depends on the accuracies of both the meteorological and the fire components of the model. An unrealistic wind forecast can quickly lead to erroneous fire spread estimates even if the fire model itself provides a perfect forecast of the fire spread. Likewise a perfect weather forecast can lead to erroneous fire spread prediction due to the inaccuracies of the fire model or the fuel data. The biases of these two models may combine, leading to drastically unrealistic results when, for example, both wind and ROS as a function of wind are overestimated. The biases may also combine to compensate, when, for example, the fire model overestimates the rate of spread but the atmospheric model underestimates the wind speed. In this study this problem is dealt with by the separate analyses of the weather and fire forecast, described in following sections.

### 4.1. Meteorological forecast

The wind field that controls fire propagation speed and direction is the three-dimensional time-varying coupled atmosphere-fire wind field, and wind is the primary meteorological factor affecting fire spread as simulated by WRF-Sfire. An evaluation of the fire spread forecast by WRF-Sfire starts therefore with an evaluation of the wind forecast by WRF.

There were 15 automated meteorological stations located within D04, the fire model domain. Unfortunately, because of very strong Santa Ana winds and the fire, the operations in many stations were disrupted. Due to missing data and data quality problems, 8 stations were dropped from the analysis. From the 7 remaining stations, four were selected, two in the center of the domain (GOSC1 and KRNM), one at the eastern boundary (PIHC1), and one at the northern model boundary (VLCC1), for analysis. The locations of these stations as well as origins of the Witch and Guejito fires are presented in Figure 2.

The resolution of the WRF simulation affects to what degree the model resolves local topography. The elevation data as incorporated into WRF are an approximation of the real topography. Figure 3 compares the elevation data interpolated to the atmospheric grid (500m resolution) to the elevation data interpolated to the fire grid (20m resolution). Therefore the actual elevation of the meteorological reporting stations and the elevation of the model terrain at these locations differ generally by up to 20m. This bias in the WRF elevation must be considered when converting the model 10m wind values to the 6.1m height used for wind measurements at meteorological reporting stations. The simple power law formula by Sedepian (1980) is applied:

$$WS_{6.1m} = WS_{10m} \times \left( \frac{6.1}{10 + (HGT_{WRF} - HGT_{ST})} \right)^{1/7},$$

where $WS_{6.1m}$ is the WRF-simulated wind speed adjusted to 6.1m, $WS_{10m}$ is the 10m wind speed from model output, $HGT_{WRF}$ is the elevation of station location on model grid, and $HGT_{ST}$ is the true elevation of the meteorological station. Eq (1) accounts for the total height bias between the model and the station due to the elevation mismatch ($HGT_{WRF}$-$HGT_{ST}$) and the difference between the station reporting height (6.1m) and the model wind output height (10m).

The WRF-Sfire simulation started on October 21$^{st}$ at 12:00 UTC and it was run for the period of 72 hours without updating the model state with current meteorological observations during the run. This means that, unlike an actual operational forecast during which the WRF state is automatically updated as the most current meteorological observations are assimilated, we do not interfere with the model as it runs. We provide the model with boundary conditions only at 12:00 UTC; i.e., the WRF-Sfire model is not nudged towards the current observed state.

The time series of the forecasted and observed wind speeds and wind directions at the four meteorological stations are presented in Figure 5. According to meteorological convention, the wind direction is the direction the wind is blowing from, represented as an angle from the north: 0 degrees means wind blowing from the north; 90 degrees means from the east; 180 degrees means from the south; and 270 degrees means from the west.

Figure 5 shows the initially weak to moderate wind increasing very quickly as the Santa Ana event starts. At 8h into the simulation, stations report wind speeds reaching approximately 14 to 20 m/s (31 to 45 mph or 50 to 72 km/h). The model captured this initial stage fairly well. The Valley Center (VLCC1) station (Figure 5c) shows a delay and underestimation in the WRF forecasted wind speed. It is possible that the WRF wind forecast at this particular station was affected either by the topography smoothed out at the model boundary or by the too slow westward simulated propagation of the Santa Ana event. The fact that the arrival of Santa Ana winds at the central stations (GOSC1 and KRNM) was captured correctly suggests that the former hypothesis is more probable.

The Witch Fire was ignited in the Witch Creek area east of Ramona, California, almost exactly 8 hours into the simulation when the wind speed picked up. A violent easterly wind caused electrical power-line arching that ignited the fire (Maranghides and Mell 2011). At that time, the Pine Hill station (PIHC1; Figure 2) reports wind speeds of 20 m/s (~45 mph or 72 km/hr). Figure 5a shows that simulated wind speeds near the ignition location were also near 20 m/s. Note that even though the local wind speed is captured very well at time of ignition, the model had some problems capturing local wind directions. As shown in Figure 5a, the forecasted wind direction was almost steady and easterly, while the Pine Hill (PIHC1) station reports variable wind directions. The wind had a northerly component at PIHC1, while the three other stations report mainly easterly winds with only a slight northerly component.

The Guejito Fire started twelve and a half hours later at 1:00 am Monday, October 22, 2007 at Guejito Creek drainage, on the South Side of California State Route 78 and 0.4 km (1/4 mile) west of Bandy Canyon Rd., or 10 km (6 miles) northeast of The Trails. The cause of ignition was identified as energized power lines contacting a lashing wire (Maranghides and Mell 2011).

After a slight decrease in the wind speed observed roughly 12h into the simulation, the Santa Ana winds strengthened again, reaching almost 25 m/s (56 mph or 90 km/hr) 12h later. Figure 5 shows that the model captures this drop and increase in the wind speed relatively well, but this varies among the analyzed locations. At the Goose Valley (GOSC1) station (Figure 5b), the model forecasts the wind directions correctly but sometimes overestimates wind speed by about 5 m/s. Fortunately the fire front is still a ways east at that time, where the wind speed forecast is significantly better (see Figure 5a). It is therefore likely that this discrepancy in the wind forecasts did not impair the fire spread forecast significantly for this time period.

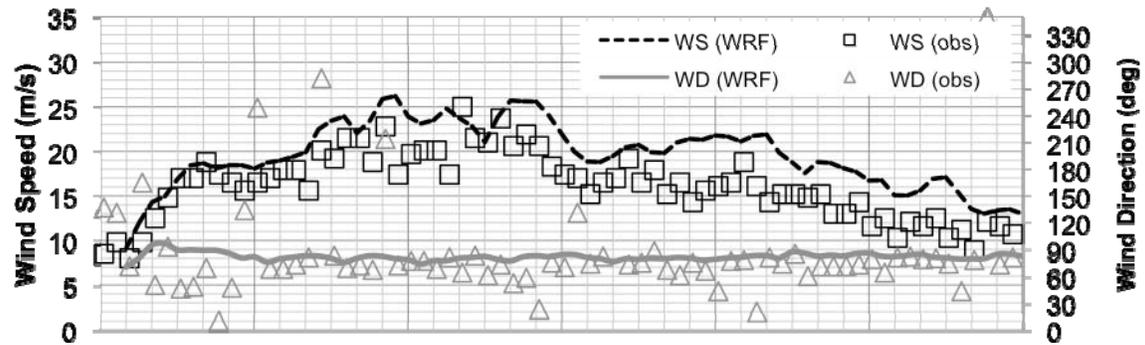
b) Goose Valley (GOSC1) station

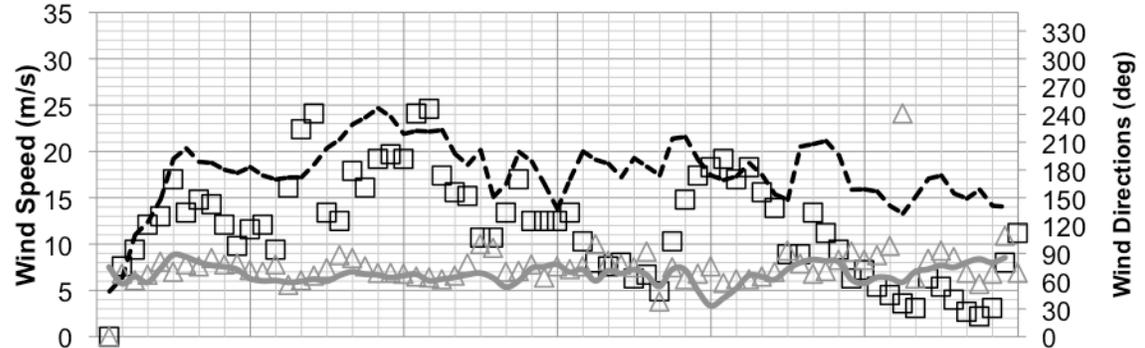
c) Valley Center (VLCC1) station

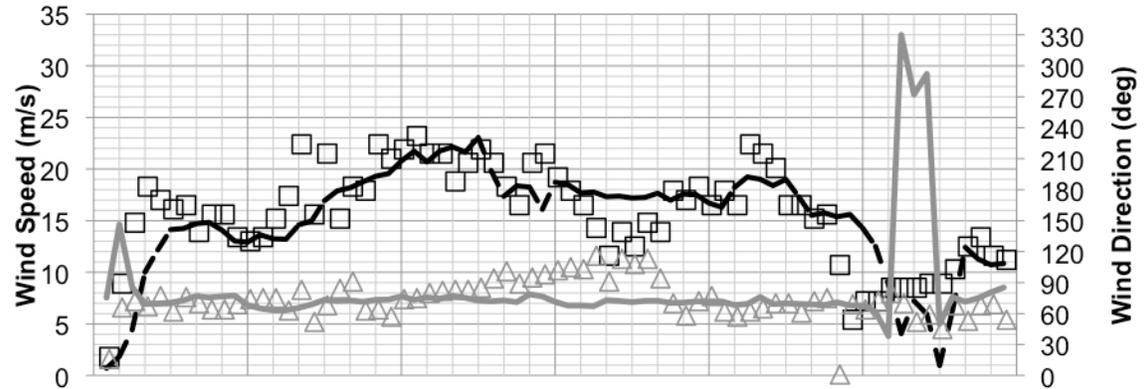
d) Ramona Airport (KRNM) station

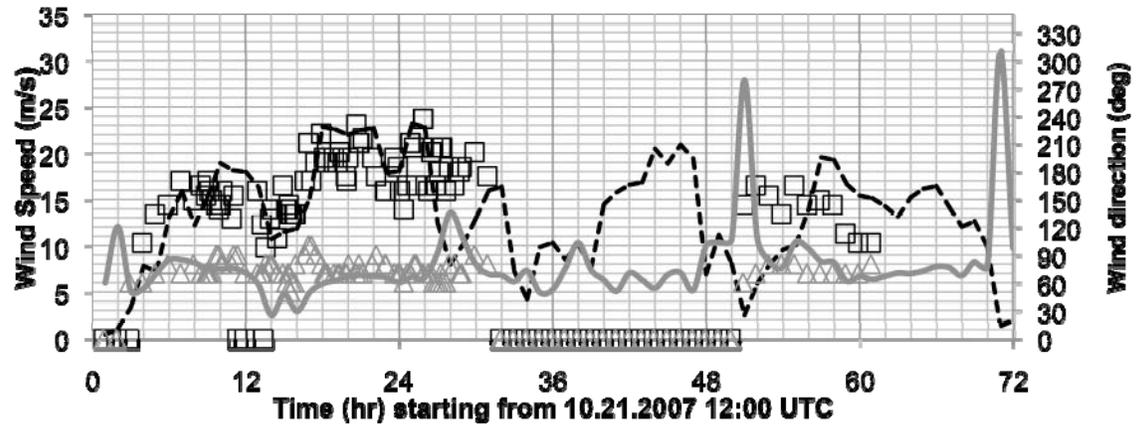

Figure 5. Time series of WRF-simulated and observed wind speed (WS) and wind direction (WD).

During the next 24h (up to 48h into simulation) the wind speed gradually decreased (Figure 5). The model generally captured this trend. There are however some discrepancies. A comparison between observations and model results at Pine Hill, the eastern-most station in the fire domain, shows that after 36h into the simulation there is a positive bias in the model results (Figure 5a). For Goose Valley (Figure 5b) the modeled wind speed is overestimated as well, but not consistently as in the case of the Pine Hill station. The Goose Valley station reported significantly higher wind speed variability compared to the other three stations. This station is located in a narrow (roughly 1km) valley surrounded by the mountains from west, east and south (see topography presented in Figure 3). Therefore, it is possible that the small-scale flow features including down slope and canyon winds induced by the complex topography of this region were not captured correctly, and led to observed discrepancies between the simulated and observed wind speed.

For the Valley Center (Figure 5c) the forecasted wind speed and direction remain in a good agreement with observations for the most of the simulated period. However, between 36 and 48h into simulation there is a period of noticeable bias in the forecasted wind speed direction. During this period, the forecasted wind speed stayed between 17 and 18 m/s while the VLCC1 station recorded a drop from 18 m/s to 12 m/s, followed by a rise back to 18 m/s. This decrease in the wind speed was associated with a change in the wind direction from ENE to ESE, which was completely missed by the model. It is noteworthy that this station is located just 2.8km (5 grid points) south from the northern domain boundary, in a region directly affected by the outer domain (D03). This distance is too short to allow the inner domain to resolve a local flow. The first 5 grid points (relaxation zone) are used to blend the outer domain forcing into the inner domain. Therefore, at that distance from the boundary the forcing coming from the outer domain may dominate, deteriorating the results near the boarder of the inner one.

Due to the evident technical problems of the Ramona Airport station, the model results between 32h and 50h cannot be validated for this location. Nonetheless, during the period of first 30 hours for which data are available the simulated wind speed and direction match observations pretty well.

### 4.2. Fire spread forecast

Due to the limited availability of fire progression data, a validation of the simulated wildfire spread is often more difficult than the validation of a meteorological forecast. Meteorological data are normally available from automated weather stations at 1-hour intervals at least. Fire data contain normally only the final fire perimeter that is, in most cases, the product of a post-fire analysis. Detailed information on fire progression is usually not recorded.  The final burnt area is affected to some degree by suppression activities that are not precisely quantified and cannot be taken into account by current fire spread and behavior models. This poses a serious limitation when using a final fire perimeter for validation of wildfire prediction models.  Nonetheless, we use whatever information is available to validate the WRF-Sfire fire progression and fire perimeter as best we can.

### 4.2.1. Fire progression

To reconstruct the propagation of the Witch and Guejito fires, reports prepared by California Department of Forestry and Fire Protection (CalFire) for incidents 07-CA-MVU-10432 (Witch fire) and CA-MVU-10484 (Guejito fire) were analyzed. These reports (available from http://www.fire.ca.gov/fire_protection/downloads/redsheet/) provide estimates of time and location of ignitions. Witness testimonies contained in these reports help to estimate the progression of the Witch and Guejito fires. Based on these testimonies three distinct locations at which the fires were observed were selected: Guejito bridge on HWY 78 at the origin of the Guejito fire; the cross-section between the Eagles Crest Road and Highland Valley Road; and N-W boundary of Rancho Bernardo. These three locations are marked on Figure 6 as a red circle, orange triangle, and yellow square, respectively. The reported timings of fire arrival at these locations were used for the validation of the simulated fire spread.

The Witch fire was ignited about 3km south-west from Santa Ysabel (see Figure 8 b), and, driven by strong, easterly 20 m/s Santa Ana winds, propagated very quickly towards the south-west. Figure 6a shows during the first 6h, the Witch fire front advancing by almost 17km, reaching a ROS of 0.79 m/s (2.8 km/h). During the next 6h, it expanded in the N-S direction, surrounded residential areas of Ramona, and extended rapidly toward Eagle Crest / Highland Road, position marked as orange triangle on Figure 6b. According to fire reports, the Witch fire reached this location at 13:00 UTC, roughly 2 hours before it was observed at the Guejito bridge (red circle on Figure 6). This time difference indicates that the fire propagated south from the origin of the Guejito fire and Guejito bridge. The numerical simulation agrees with this scenario. The simulated fire left the origin of the Guejito fire and advanced S-W, around 4km to the N-W, and reached Eagle-Crest / Highland drive at 12:30 UTC (30 minutes earlier than observed).

The Geujito fire ignited at 8:00 UTC (20h from the start of the simulation) and started its quick westward propagation. In 5h and 15minutes it advanced southward by more than 8 km. The moment and location of merging Witch and Guejito fires seem to be captured correctly. In the simulation the Witch merged with Guejito fire at the Guejito bridge at 15:00 UTC, while the reported time of Witch arrival at this location was 14:45 UTC. One of the witnesses, whose property was destroyed, reported that Guejito burnt the north-eastern part of his land, and a couple of hours later the Witch fire advancing from south-east burnt its rest. Simulated fire behavior is consistent with this testimony. Figure 6c and d indicate that the area south from the Guejito creek was affected initially by the Guejito fire and then by the Witch fire approaching from S-E.

After merging, both fires continued to advance toward San Bernardo, increasing their N-S extent (see Figure 6e). The Santa Ana system advanced from East, gradually weakening as it approached the coast. Therefore the wind speed over the western part of the fire domain was significantly smaller than it was in the center and eastern parts. That may explain the significant fire deceleration evident in Figure 6 panels d) and e).

Records of firefighters' interventions in San Bernardo suggest that fire reached its N-W boundary around 18:30-19:30UTC, roughly 6h after burning properties on its N-E edge.

Even though WRF-Sfire is not able to simulate structural fires, it was able to simulate the wildfire arrival at these locations with great accuracy. According to the WRF-Sfire forecast, the fire reached the eastern side of San Bernardo at 12:30UTC, then went around it and reached its western side at 18:50UTC. Figure 6c shows the simulated fire perimeter at 13:15 UTC, soon after 12:30 UTC.

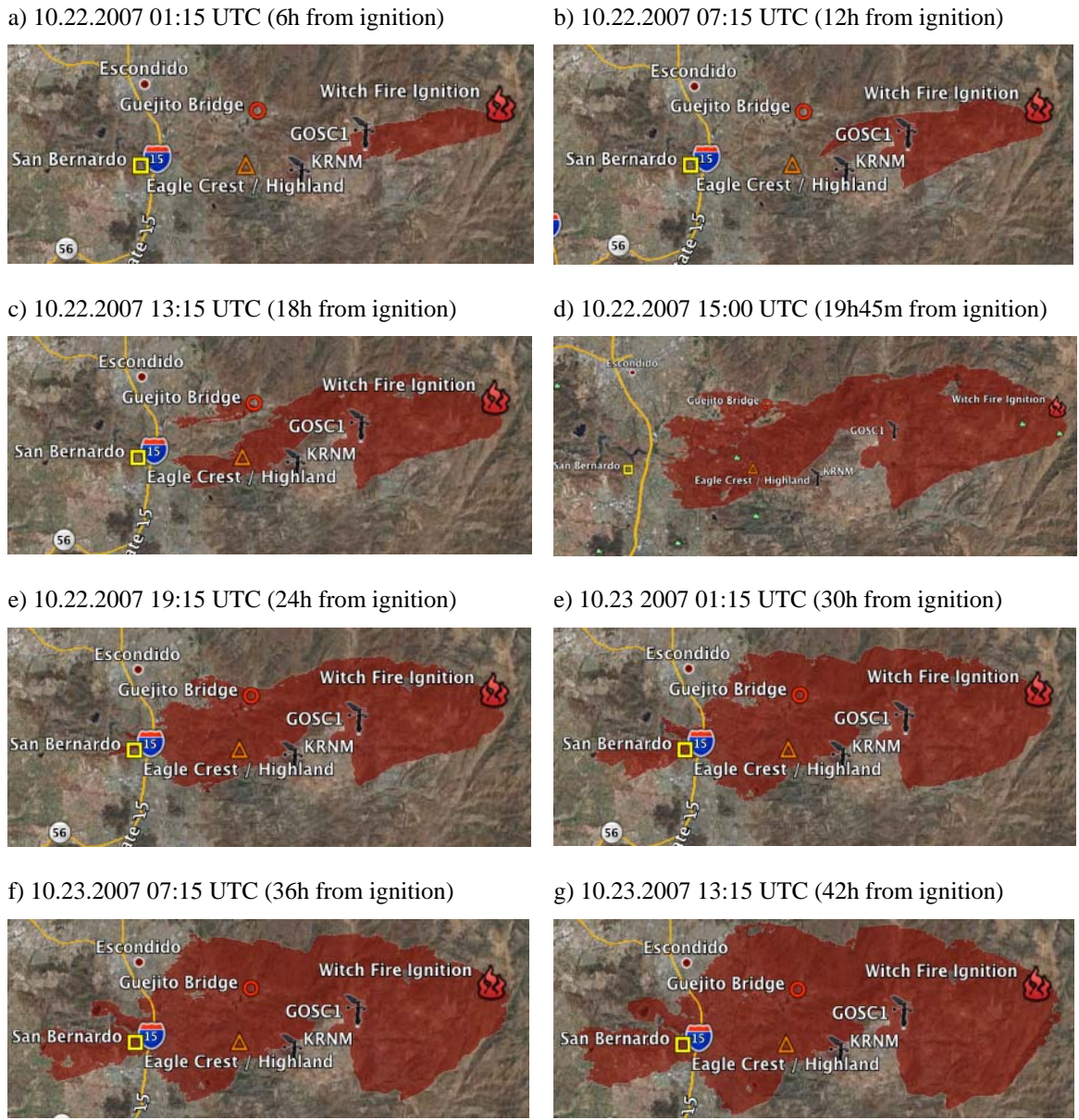

Figure 6. Simulated progression of Witch and Guejito fires. Image overlay prepared using Google Earth.

Figure 6 panels e), f) and g) show further fire progression. The E-W extent of the fire did not increase significantly after 1:15 UTC, but its N-S spread continued. Note that the fire, after consuming most of the fast-burning grassy fuels types (yellow and orange colors on fuel map shown in Figure 4), reached an area with less combustible fuel types and slowed down. The last two panels of Figure 6 show that the fire area did not increase substantially between 7:15 and 13:15 UTC.

### 4.2.2. Burnt area and fire perimeter

A wildland fire will stop spreading from lack of combustible fuel and ignition temperatures, and changing wind conditions. There is no definite point when a WRF-Sfire simulated fire stops. Therefore, it is necessary to define a time when the model fire reaches its "final" perimeter. Figure 7 shows the total burnt area and change in equivalent fire diameter at 10-minute intervals for the simulation. This plot is used to identify the moments when the WRF simulated area matches the observed one and when the model fire "stops" propagating.

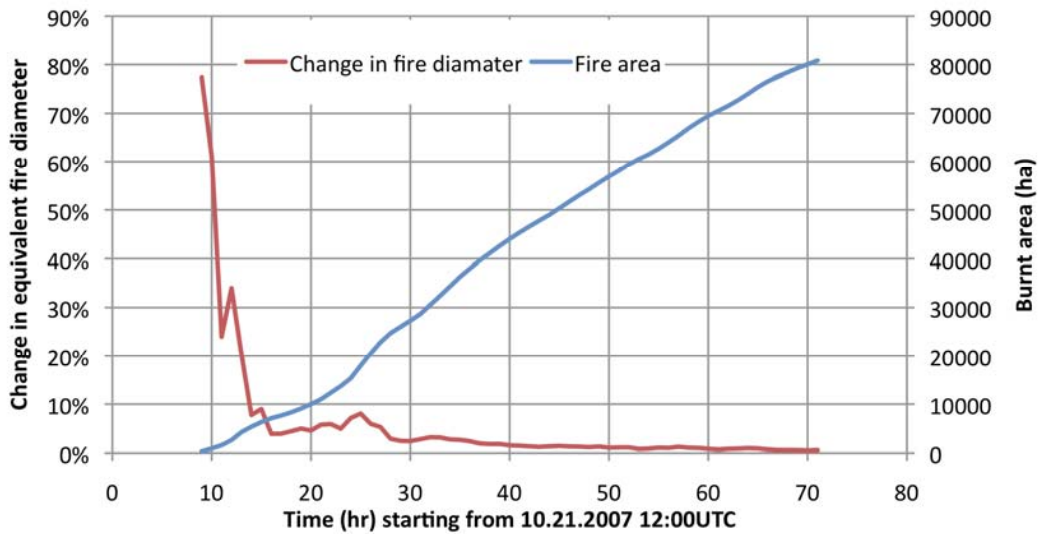

Figure 7. Time series of the burnt area and the relative change in the equivalent fire diameter.

Initially we planned to use 56,796 ha, Keeley (2009)'s estimate of total area burnt by the Witch and Guejito fires, to determine the model fire's final perimeter. However since Keeley's area estimate subtracted the unburned area within the fire perimeter, we decided that a different estimate of the total area encompassed by the final fire perimeter should be used. The reason Keeley's area estimate is not appropriate for model validation is that in WRF-Sfire there is no mechanism to stop the fire, other than a lack of fuel or fuel moisture above the fire extinction threshold. Once ignited, the available fuel is burned and as a result the WRF-Sfire simulated final fire area is always uniform, with unburned patches corresponding only to incombustible areas. Therefore, in addition to the remotely-sensed burnt area, we determined a total area encompassed by the final fire perimeter that included unburned areas within the observed final fire perimeter. This is estimated as 70,817 ha. As shown in Figure 7, the simulated fire area reached this value 60 hr into the simulation (24.10.2007 00:00 UTC), while the remotely-sensed area of 56,796 ha was reached around 48h into the simulation (23.10.2007 12:00 UTC). These two moments are used for validation of the WRF-Sfire simulated fire area presented in Figure 8.

Additionally, we also computed as a function of time the equivalent fire diameter as the diameter of a circle having the same area as the WRF-Sfire. This is the red line plotted in Figure 7. We used its relative change as a measure of the rate of fire perimeter growth. Figure 7 shows that the relative changes in the equivalent fire perimeters are very low for the times where the simulated fire area matched the observed one; i.e., 1.14% at 48h and 0.82% at 60h into the simulation. This suggests that, even though the simulated fire didn't stop, the 1% change in the relative diameter could be used as a threshold for defining the end of its active spread.

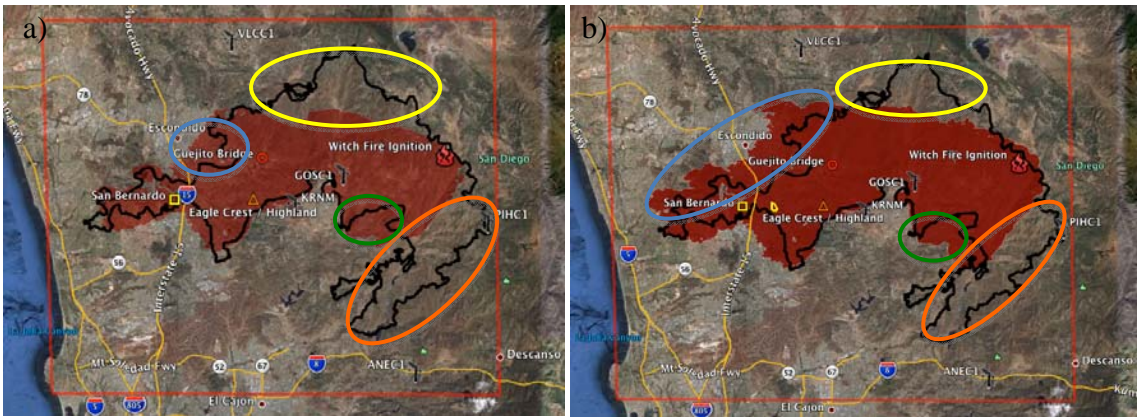

Figure 8. Observed final fire perimeter (black contour) vs. WRF-Sfire simulated fire area (red fill) for: a) 23.10.2007 12:00 UTC (48h into simulation); and b) 24.10.2007 00:00 UTC (60h into simulation).

Figure 8 shows that the general shape of the fire perimeter was captured well in some regions and not in others. The four main areas where discrepancies between simulated and observed fire perimeters occur are marked in Figure 8 using color ovals. The overall difference between the actual and simulated final fire perimeters is in the N-S extent. The simulated fire did not reach as far north (see yellow ovals in Figure 8) as the observed final fire perimeter suggests. The elevation between the northern edge of the simulated fire and the northern edge of the observed final perimeter changes by roughly 350m over a distance of 7km. This inclination (below 3 degrees) is not large enough to generate any measurable upslope ROS. This suggests that it was a southerly wind component that drove the fire perimeter north. The wind speed measured at the VLCC1 station (see Figure 5 c) indicates that there was a period of southeasterly wind (wind direction around 110 deg) between 36 and 48h. The model did not capture this shift in the wind direction. Figure 5c shows that during this period the model predicted a wind direction of 80 degrees and higher wind speeds than observed. There is no guarantee that the wind speed conditions recorded by the VLCC1 station can be treated as representative for the disagreement area we consider. Nonetheless, if the measured southerly wind component of 4.4 m/s was supplied to the Rothermel model, the WRF-Sfire fire could have advanced quickly over the grassy fuel by about 8 km in that time, which is exactly the width of the gap between the observed and simulated fire perimeter.

Another region where the simulated fire perimeter does not match the observed final perimeter is along the south-eastern edge of the fire marked by the orange oval in Figure 8. The wind speed and direction time series from the nearby PIHC1 station (Figure 5 a) also show slight discrepancies between the simulated and measured wind directions that potentially could have contributed to this mismatch. Between 12h and 24h into simulation, during the period of increasing wind speed, there were a few reports of winds with a westerly component. However, since these shifts in the wind direction were only temporary, they cannot be treated as the only reason for observed discrepancies. It is more probable that the ROS on the flanks of the simulated fire was generally too low, leading to too narrow a fire spread in N-S direction. A test simulation performed (not shown) with an increased default ROS (i.e., 0.1 m/s as opposed to 0.0225 m/s) for grassy fuels showed significant improvement in the shape of the south-eastern edge of the model fire. That suggests that the underestimation of the model's default ROS on the fire flanks spoiled the results.

The blue and green ovals in Figure 8 highlight regions that were burnt in the WRF-Sfire simulation but not in reality. The blue oval shows the residential area of Escondido and the green one corresponds to San Diego Estates. The fuel in these two areas was not marked as incombustible, so the model fire advanced through them.

The Guejito and Witch fires burned for 10 days. During this period various suppression actions were taken that probably affected the fire shape, but that are not purposely accounted for in the simulation. Also, the simulation ends much earlier (roughly 65h after Witch fire ignition) than the time the final fire perimeter was reconstructed from post-fire operations. Precise data showing which parts burned during the initial 3-day Santa Ana event and which burned during the last 7 days of the fire are not available, making it impossible to assess unambiguously the agreement between the progression and final position of the simulated and observed fire perimeters. However, the fire spread data seem to confirm that the fire progression during the Santa Ana winds was captured realistically in the simulation, with only a 30 minute mismatch between model fire arrival times and the observed fire arrival times.

## 5. Summary and conclusions

This pilot study demonstrates that it is possible to use WRF-Sfire, a physics-based fluid-dynamical deterministic modeling system, to provide a numerical forecast of wildfire behavior and spread in a landscape setting in real time. The entire 72h forecast analyzed in this study was computed within 4h 48min, while the first 24h forecast was available in just 1h 35min. This computational performance proves that it is feasible to use the coupled WRF-Sfire atmosphere-fire simulation for real-time wildfire forecasting.

As discussed in the Introduction, the primary physical advantage of the WRF-Sfire is that the WRF-Sfire's physics-based fluid-dynamical approach to wildfire forecasting includes the ability to model wind and coupled atmosphere/wildfire interactions. The spatially-gridded GIS data on fuels and topography, along with meteorological data from the national network of weather observations, can be easily ingested using the operational

WRF pre-processing system (WPS). The same system allows for feeding the model with 'future' boundary conditions extracted from larger-scale operational numerical forecasts. These features make the WRF-Sfire's a weather- and fire-forecasting model suitable for real world simulations. Although not done in the Santa Ana wildland WRF-Sfire simulation in this study, meteorological data from the national network of weather observations collected after the model start can be used by the operational WRF-DA data assimilation system to improve a WRF-Sfire forecast. Once a WRF weather forecast at the relatively coarse (operational) resolution is made, the system's nested-grid capabilities can provide a real-time forecast of velocity, temperature, and moisture fields at the fine resolution of the fire domain.

Simulation of wind, temperature, and moisture in the fire domain is sensitive to the lateral boundary conditions at its horizontal border. If these boundary conditions are not accurate, the simulated fields in the fire domain deteriorate. There is also a margin within each nested domain that is used to blend the lower resolution data from outer domain into the higher resolution (nested) domain. Within and close to this zone, the results should be treated with some skepticism, since they are strongly affected by the coarser outer domain that is not capable of resolving fine features expected to be seen in the finer, nested domain. We suspect that this could be a reason for some of the discrepancies between the simulated and observed wind at the VLCC1 station, located just at the northern border of the fire domain.

As shown in Section 4.1, WRF prediction of high wind speeds was done well, while prediction of weak winds especially at lower elevations and in mountain valleys was slightly worse. Comparisons to available observations indicate that the magnitudes of WRF forecasted weak and gusty low-elevation winds and down-slope or lee-slope winds were generally overestimated. Even though the wind direction was simulated well, there were also some intermittent discrepancies between simulated and observed wind speeds at PIHC1, the most northern meteorological station in the fire domain, and to smaller degree at VLCC1, the most eastern station (see periods 12-24h and 36-48h into simulation in Figure 5 a and c).

The accuracy of the WRF-Sfire for operational use was judged based on a comparison between observed and forecasted fire progression as well observed and simulated final fire perimeters. Agreement between the observed and simulated fire perimeters was good in many areas and poor in others. The northern fire extent was not captured particularly well. The most likely reason for a poor final fire perimeter forecast in the northern part are the periods of poor wind direction forecasted by WRF. The smaller than observed fire extent on the fire's S-E edge also seems to be affected to some degree by intermittent errors in the simulated wind direction. Nonetheless, the fire spread data for the center part of domain confirm that the most active progression of the fire perimeter during the Santa Ana winds was captured, with the mismatch between the simulated and observed timing of the fire arrival of 30 minutes or less.

It is noted that the modeled fire perimeter is very sensitive to the wind forecast. When the Rothermel formula moves the fire a certain distance ahead based on the WRF-Sfire winds, it can never move the fire back to the previous location. In other words, forecast errors in simulated ROS accumulate in time. Therefore even a relatively accurate wind forecast may be not good enough to provide equally accurate fire perimeter forecast.

Another source of error may be the static fuel moisture content used in the study. As reported by Jolly (2007), the Rothermel spread-rate formula is very sensitive to changes in fuel moisture content that are hard to estimate precisely. In this study the moisture content was set to a constant 6.5%. Since the actual air moisture content was lower inland than at the coast, this could cause an overestimation in the fuel moisture content over the center and eastern parts of the fire domain, and an underestimation over its western part. It may be that the fuel moisture underestimation over the western regions is manifested in overestimation in the fire spread at its western edge (see Figure 8b). Furthermore, it is likely that spatial and temporal variability of the fuel moisture throughout the entire simulation would provide a more accurate forecast of fire propagation.

The Rothermel fire spread formula was developed for a head fire, not for head-fire spread for cross-slope winds or for non-head fire spread along the fire flanks or behind a fire. When there is no flow perpendicular to the fire perimeter, the WRF-Sfire defaults to a constant no-wind ROS. It is evident that there are significant differences between observed and simulated fire spread on the flanks, and we believe that this simplified approach to ROS can be responsible for that. For this reason, another modeling technique, instead of defaulting to a constant ROS, is recommended for these wind and fire-perimeter geometries.

Another source of error in the simulation of Santa Ana fires may be the impact of sloping terrain on fire spread. The Rothermel formula used in WRF-Sfire has a slope correction factor that provides upslope fire spread in the wind direction normal to the local fire perimeter when the slope incline is greater than zero. At the same time, the WRF-Sfire wind was modified by terrain, especially by steep terrain. In many circumstances, steep terrain is a source of energy for the wind that accelerates upslope flows. In the simulation of the Santa Ana fires, the fire perimeters propagated mostly downwind so we probably don't have a double-counting problem here. In general, this may be an issue.

Even though uncertainties in wind and moisture forecasts may cancel out to provide a forecast that resembles the observations, they may also sum and spoil the results. One way to address this is to produce an ensemble forecast (Finney et al. 2011). An ensemble forecast would quantify the inherent uncertainty in WRF weather and WRF-Sfire fire spread predictions. By quantifying the uncertainty, a confidence level in the forecast is demonstrated, and the results are more useful to wildfire management than would be otherwise.

Despite the discrepancies between observed and simulated final fire spread, the results of the study indicate that the potential for operational application of WRF-Sfire is promising. More accurate or confident fire behavior and propagation prediction by WRF-Sfire has many possible uses for the wildfire management community. These can include: wildland fire evacuation planning; effective and safe deployment of aerial and ground resources; predicting wildfire and prescribed fire intensity/severity that may vary under changing local meteorological and terrain conditions; where and how to fight wildfire, for example, to prevent wildland-urban interface fires or when attempting to control wildfire in ecosystems that need protection from smoke or are at risk of severe fire damage.

**Acknowledgements**

This research was supported by the National Institute of Standards and Technology (NIST), Fire Research Grants Program, Grant 60NANB10D225. A gratis grant of computer time from the Center for High Performance Computing, University of Utah, is gratefully acknowledged.